\documentclass[runningheads,a4paper]{llncs}

\usepackage{amsmath}
\usepackage{amssymb}
\usepackage{graphicx}

\usepackage{url}
\urldef{\mailsa}\path|{lhe4, lmurp14}@u.rochester.edu, jiebo.luo@gmail.com|      
\newcommand{\keywords}[1]{\par\addvspace\baselineskip
\noindent\keywordname\enspace\ignorespaces#1}

\begin{document}

\mainmatter  

\title{Using Social Media to Promote\\ STEM Education: Matching College\\ Students with Role Models}

\titlerunning{Using Social Media to Promote STEM Education}

%
%
\author{Ling He
\and Lee Murphy\and Jiebo Luo}
\authorrunning{Ling He, Lee Murphy, and Jiebo Luo}

\institute{University of Rochester, Goergen Institute for Data Science, Rochester, NY 14627\\
{\mailsa}\\
}

\toctitle{Using Social Media to Promote STEM Education: Matching College Students with Role Models}
\tocauthor{Ling He, Lee Murphy, and Jiebo Luo}
\maketitle

\begin{abstract} 
STEM (Science, Technology, Engineering, and Mathematics) fields have become increasingly central to U.S. economic competitiveness and growth. The shortage in the STEM workforce has brought promoting STEM education upfront. The rapid growth of social media usage provides a unique opportunity to predict users' real-life identities and interests from online texts and photos. In this paper, we propose an innovative approach by leveraging social media to promote STEM education: matching Twitter college student users with diverse LinkedIn STEM professionals using a ranking algorithm based on the similarities of their demographics and interests. We share the belief that increasing STEM presence in the form of introducing career role models who share similar interests and demographics will inspire students to develop interests in STEM related fields and emulate their models. Our evaluation on 2,000 real college students demonstrated the accuracy of our ranking algorithm. We also design a novel implementation that recommends matched role models to the students. 

\end{abstract}

\section{Introduction}

The importance of the STEM industry to the development of our nation cannot be understated. As the world becomes more technology-oriented, there is a necessity for a continued increase in the STEM workforce. However, the U.S. has been experiencing the opposite. In the United States, 200,000 engineering positions go unfilled every year, largely due to the fact that only about 60,000 students are graduating with STEM degrees in the United States annually [17]. Another obvious indication is the relatively fast growth in wages in most STEM-oriented occupations: for computer workers alone, there are around 40,000 computer science bachelor’s degree earners each year, but roughly 4 million job vacancies [29]. Therefore, our motivation is to solve this problem of STEM workforce shortage by promoting STEM education and careers to college students so as so to increase the number of people who are interested in pursuing STEM majors in college or STEM careers after graduation. 

In this paper, we present an innovative approach to promote STEM education and careers using social media in the form of introducing STEM role models to college students. We chose college students as our target population since they are at a life stage where role models are important and may influence their career decision-making [15]. Social media is useful for our study in the following two ways: 1) the massive amount of personal data on social media enables us to predict users’ real life identities and interests so we can identify college students and role models from mainstream social networking websites such as the microblogging website Twitter and professional networking website LinkedIn; 2) social media itself also can serve as a {\it natural and effective platform} by which we can connect students with people already in STEM industries. 
\begin{figure}
\vskip -0.1in
\centering
\includegraphics[height=6.2cm,width=\linewidth,keepaspectratio]{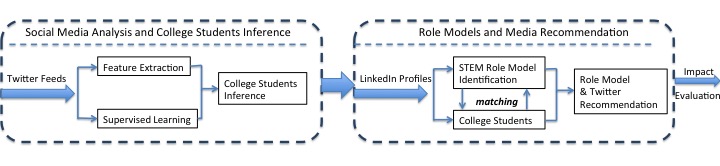}
\caption{The framework for promoting STEM education and careers using social media to match college students with STEM role models.}
\label{fig:example}
\end{figure}

Our approach is effective in the following three ways. First, increasing STEM presence will inspire students to develop interests in STEM fields [18]. Second, the exposure of career STEM role models that students can identify with will have positive influence on students, as strongly supported by previous studies [12]. Finally, as a form of altruism, accomplished people are likely to help young people [11,6] and people who resemble them when they were young [21]. More importantly, social learning theory [1,2], psychological studies, and empirical research have suggested that students prefer to have role models whose race and gender are the same as their own [12,30,15] as well as who share similar demographics [7] and interests [16]. Motivated and supported by the findings of these related studies, we select {\it gender, race, geographic location, and interests} as the four attributes that we will use for matching the students with STEM role models. In addition, similar interests and close location will further facilitate the potential {\it personal connection} between the students and role models. 

In particular, we first use social media as a tool to identify college students and STEM role models using the data mined from Twitter and LinkedIn. As a popular online network, on the average, Twitter has over 350,000 tweets sent per minute [27]. Moreover, in 2014, 37\% social media users within the age range of 18-29 use Twitter [5]. This suggests a large population of college users on Twitter. In contrast, as world’s largest professional network, LinkedIn only has roughly 10\% college users out of  more than 400 million members [25], but has a rich population of professional users. Part of its mission is to connect the world's professionals and provide a platform to get access to people and insights that help its users [14]. Our goal, to connect college students with role models, is {\it organically consistent with LinkedIn's mission and business model}. 

Specifically, we train a reliable classifier to identify college student users on Twitter, and we build a program that finds STEM role models on LinkedIn. We employ various methods to extract gender, race, geographic location and interests from college students and STEM role models based on their respective social media public profiles and feeds. We then develop a ranking algorithm that ranks the top-5 STEM role models for each college student based on the similarities of their attributes. We evaluated our ranking algorithm on 2,000 college students from the 297 most populated cities in the United States, and our results have shown that around half of the students are correctly matched with at least one STEM role model from the same city. If we expand our geographic location standard to the state-level, this percentage increases by 13\%; if we look at the college students who are from the top 10 cities that our STEM role models come from separately, this percentage increases by 33\%. 

Our objective is to do social good, and we expect to promote STEM education and careers to real and diverse student population. In order to make a real life impact on the college students after we obtain the matches from the ranking algorithm, we design an implementation to help establish connections between the students and STEM role models using social media as the platform. For each student, we generate a personalized webpage with his top-5 ranked STEM role models' LinkedIn public profile links as well as a feedback survey, and recommend the webpage to the student via Twitter. Ultimately, it is entirely up to the student and the role models if they would like to get connected via LinkedIn or other ways, and we believe these connections are beneficial for increasing interest in STEM fields. It is noteworthy that {\it LinkedIn has already implemented a suite of mechanisms to make connection recommendations}, even though none of which is intended to promote STEM career specifically. Fig.2 illustrates how our implementation naturally fits into the work flow and business model of LinkedIn. 
\begin{figure}
\vskip -0.2in
\centering
\includegraphics[height=6.2cm,width=\linewidth,keepaspectratio]{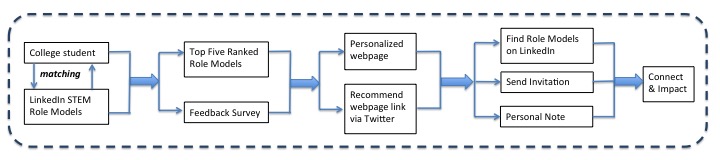}
\caption{The framework of our implementation to help establish the connections between the college students and the STEM role models.}
\vskip -0.1in
\end{figure}

Our study has many advantages. Leveraging existing social media ensures that we are able to retrieve a large scale of sampling users and thus our implementation is able to influence a large scale of students. Also, due to available APIs and existing social media infrastructures, our data collection and our implementation are low cost or virtually free. More importantly, unlike some traditional intervention methods, we recommend STEM role models to college students in a {\it non-intrusive} way. We tweet at a student with the link of his personalized webpage, and it depends on himself if he wants to take actions afterwards. Finally, our approach is failure-safe in delivery. If there are some Twitter users that are classified incorrectly as college students, it has no harmful impact on them even if we promote STEM education to them. 

The major contributions made in this study are fourfold. First, we take advantage of social media to do social good in solving a problem of paramount national interest. Second, we take advantage of human psychology, motivation, and altruism. That people are more likely to be inspired by models who are like them, and people who are accomplished are likely to help young people who share similarities with them. Third, we have developed a simple yet effective ranking algorithm to achieve our goal and verified its effectiveness using real students. Lastly, we design an implementation that seamlessly mashes up with the natural work flow and business model of LinkedIn to establish the connections between students and role models. 

\section{Related Work}

STEM workforce is significant to our nation, and the shortage in such fields makes promoting STEM education and careers indispensable. We review the existing methods of promoting STEM education and build on previous research in both computer science and human psychology.

Previous effort has been made to promote STEM education. Most existing intervention methods focus on promoting through school educators [19], external STEM workshops [26], and public events such as conferences [22]. However, very little evidence has shown that these strategies were effective. On the other hand, while none of the methods has utilized the rich database and powerful networking ability of social media, social media-driven approaches have succeeded in many applications, such as health promotion and behavior change [33].

The abundance of social media data has attracted researchers from various fields. We benefit the most from studies that related to age prediction and user interest discovery. Nguyan et al. [20] studied various features for age prediction from tweets, and guided our feature selection for identifying college students. Michelson and Macskassy [31] proposed a concept-based user interest discovery approach by leveraging Wikipedia as a knowledge base while Xu, Lu, and Yang [32], Ramage, Dumais, and Liebling [23] both discovered user interest using methods that built on LDA (Latent Dirichlet Allocation)[3] or TF-IDF [24]. 

Our study also adopts knowledge from psychological studies that demonstrate the importance of having a role model with similar demographics and interests. Karunanayake [12] discussed the positive effect of having role models with the same race, and it holds across different races; Weber and Lockwood [30] discovered that female students are more likely to be inspired by female role models; Ensher and Murphy [7] indicated that liking, satisfaction, and contact with role models are higher when students perceive themselves to be more similar to them in demographics; and Lydon et al. [16] suggested that people are attracted to people who share similar interests. These studies help determine the attributes that we selected to match the students with role models. 

\section{Data}

We used the REST API to retrieve Twitter data. Instead of directly searching for college Twitter users among all the general users, we focused on the followers of 112 U.S. college Twitter accounts since there is higher percentage of college students among these users. In total, we successfully retrieved more than 90,000 followers. For each user, we extracted the entities of his most recent 200 tweets (if a user has fewer than 200 tweets, all his tweets were extracted) and his user profile information, which includes geographic location, profile photo URL, and bio. After we filtered out API failures, duplicates, and users with zero tweet or empty profile, we are left with 8,688,638 tweets from 62,445 distinct users. 

Due to the limited information that LinkedIn API allows us to retrieve, we employed web crawling techniques to obtain the desired information directly from the webpage. We built a program that does automated LinkedIn public people search and used it to search users based on the most common 1,000 surnames for Asians, Blacks, and Hispanic, and more than 5,000 common American given names \footnote{All the names were retrieved from \url{http://names.mongabay.com}}. Despite some overlapping surnames, the large number of names we searched is still able to ensure the diversity of the potential role models, and our results confirmed that. For each search, the maximum number of users returned is 25, and we collected the public profile URLs of all the returned users. After we deleted the duplicates, we retained 182,016 distinct LinkedIn users.

\section{Identifying Twitter College Student Users}

We employed machine learning techniques to identify Twitter college student users (i.e. from incoming freshmen to seniors). First, we labeled our training set. We used regular expression techniques to label college student users and non-college student users. Specifically, we studied patterns in users' tweets and bio, and constructed 45 different regular expressions for string matching. For example, expressions such as ``I'm going to college'', ``\#finalsweek'', or ``university '19'' are used to label college students; and expressions such as ``professor of'', ``manager of'', or ``father'' are used to label non-college students. If a user's tweets and bio do not contain any of the 45 expressions, the user is unlabeled. We then manually checked and only counted the correctly labeled users. In the end, we are left with 2,413 labeled users, where 1,103 are college students and 1,310 are non-college students, as well as 60,032 unlabeled users.

Second, we trained our labeled data set to develop a reliable classifier using the LIBSVM Library [4] in WEKA [8]. We chose SVM for our binary classification because it is efficient for the size of our data set. We learned from Nguyan et al.'s study of language and age in tweets [20] that the usage of emoji, hashtag, and capitalized expressions such as ``HAHA'' and ``LOL'' are good age indicators. We built on their study and took a step further to use these three features for differentiating college students (i.e. specific age group) from general users. We were also curious about whether re-tweet would be another good age indicator, so we also extracted this feature. For each user, each feature is represented by its \emph{relative frequency} among the user's tweets:
\begin{equation}
   \frac{\text{\# of tweets that contain this feature}}{ \text{total \# of tweets}}
\end{equation}
Since relative frequencies are continuous, we discretized them into 10 bins with an equal width of 0.1 and assigned them with ordinal integer values for classification.
\begin{figure}
\vskip -0.2in
\centering
\includegraphics[height=6.2cm]{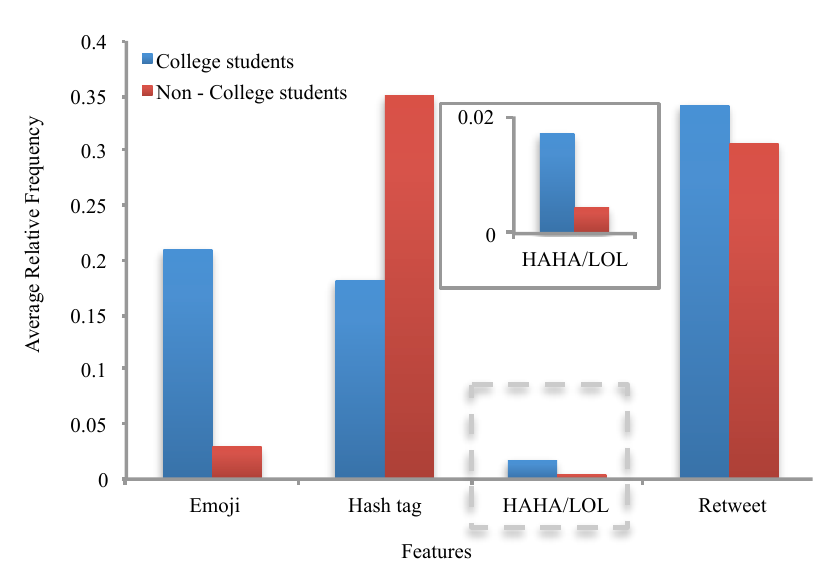}
\caption{Average usages of the four features for \emph{college} and \emph{non-college student users} among labeled Twitter users.}
\vskip -0.1in
\end{figure}

Fig.3 demonstrates our analysis of the four features. On average, college student users use emojis and HAHA/LOL more frequently while non-college student users use hashtags more frequently. We note that these results are consistent with the conclusions of a previous study [20]. However, there is not much difference in re-tweet between these two groups. We experimented training the classifier with and without re-tweet, our 10-folds cross-validation results showed that including re-tweet actually slightly lowers the accuracy of the classifier. Thus, we confirmed that re-tweet is a noise and does not help us to differentiate college student users. Our final classifier trained from the other three features achieves a high accuracy of 84\%. We then used this trained classifier to infer college student users among the unlabeled users. We further labeled 18,351 users as college students, and with our manually labeled college student users, together we have labeled 19,454 college student users in total. 

\section{Finding LinkedIn STEM Role Models}

Our goal is to find diverse STEM role models from LinkedIn in terms of geographic locations and industries. While the definition of a role model is subjective to an individual student, we take an objective view and consider people who have received STEM education and work in STEM-related industries or have a career in STEM industries as role models. 

We first filtered out users who are outside of the United States and then built a \emph{Role Model Identification} program to find STEM role models. The program takes in a user's profile URL, crawls the contents in ``industry'' and ``education'' fields on the user's profile and only outputs the URL if the user is a STEM role model. Specifically, we divided all 147 LinkedIn industries into three groups, ``non-STEM'', ``STEM'', and ``STEM-related''. For example, ``Biotechnology'' and ``Computer Software'' are ``STEM'',  ``Music'' and ``Restaurants'' are ``Non-STEM'', and ``Financial services'' and ``Management consulting'' are ``STEM-related''. We only consider those users who are under ``STEM'' or under ``STEM-related'' with a degree in STEM majors as role models. We used the 38 STEM majors offered at our University as our standard.  
\begin{figure}
\centering
\includegraphics[height=6.2cm]{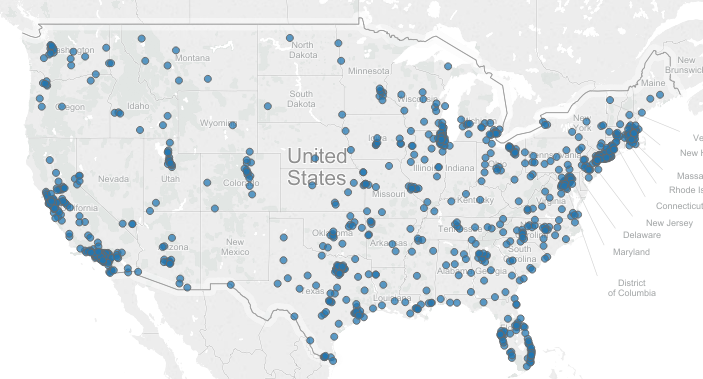}
\caption{The geographic location distribution of \emph{STEM role models}. Darker color indicates higher density.}
\end{figure}

After we obtained the profile URLs of STEM role models, we crawled their entire profiles using the URLs. We successfully found 25,637 STEM role models from 2,022 distinct locations in the United States, including some places in Hawaii and Alaska. Fig.4 shows a rough visualization of the diverse geographic locations the STEM role models come from. The top-10 cities that role models come from are, not surprisingly, San Francisco, New York City, Atlanta, Los Angeles, Dallas, Chicago, Washington D.C., Boston, Seattle and Houston.  

\section{Matching College Students With Role Models}

This section presents the methods we employed to extract the gender, race, geographic location and interests from college students and STEM role models as well as our ranking algorithm that matches them based on the similarities of these attributes. We reiterate that our selection of attributes are supported by a variety of previous related studies. These factors can make the most influential pairing because they ensure that a student gets a mentor with a similar background for affinity. Moreover, close geographic location and similar interests are valuable for potential real life interaction between the students and role models. 

\subsection{Gender and Race Extraction}

We extracted race and gender from both textual and visual features, namely the users' names and profile photos. We recognize that there are people who identify themselves with genders other than male and female; we also recognize that there are a variety of ways for categorizing races. To build a prototype system, we will use male, female for gender categorization, and use White, Black, Asian, Asian Pacific Islander (i.e. Api) and Hispanic for race categorization. 
\begin{figure}
\centering
\vskip -0.1in
\includegraphics[height=6.2cm]{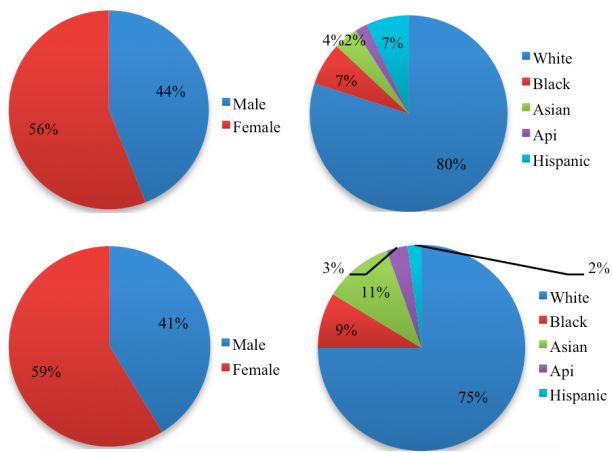}
\caption{Distribution of gender and race. Top row: college students; Bottom row: STEM role models.}
\end{figure}

In particular, we used Genderize.io \footnote{A database that contains 216,286 distinct names across 79 countries.}, Face++\footnote{A face detection service that detects 83 points of the face and analyze features such as age, gender, and race.}, and Demographics\footnote{A database that contains U.S. census for demographics.} to extract these two attributes. Genderize.io and Demographics predict gender or both gender and race based on the user’s given name or full name while Face++ predicts both using the user's profile photo. In total, we obtained three gender predictions and two race predictions for each user. Each prediction is returned with an accuracy, and in the case the tool fails to predict, the prediction will be null. We picked the gender and race predictions with the highest accuracy. 

As a result, we extracted the gender of 80\% college students and 97\% role models, and the race of 46\% college students and 92\% role models. Almost all role models have both attributes since we used their LinkedIn profiles, where the profile photos are usually high quality and the names are usually real. In contrast, Twitter profiles sometimes can contain profile photos with random objects and invented names. Fig.5 shows the make-up of those college students and STEM role models whose gender and race were successfully extracted. 

\subsection{Location and Interests Extraction}

We directly extracted geographic locations from the ``location'' field in Twitter and LinkedIn profiles. The interests extraction is less straightforward and we used other features as proxies for this attribute.

We were able to extract the locations of all STEM role models since LinkedIn requires users to have a valid geographic location on their profiles. These locations usually contain the city and the state that role models work in. However, Twitter does not have this requirement, and we noticed that not every college student has filled the location field on his profile and some of the filled locations are not valid. In fact, 34\% Twitter users either did not fill the ``location'' field or provided fake geographic locations; among those valid locations, roughly 65\% are at city-level [11]. In addition, we observed that many students use the name of their educational institutions as locations, and some locations are not correctly spelled or formatted. For example, a student's location is ``mcallentx'', which refers to the city McAllen in Texas, but not a place called ``mcallentx''. 

Due to the difference in the nature of LinkedIn and Twitter, we selected different proxies as interests for role models and college students. For role models, we directly extracted the contents in ``interests'' and ``skills'' fields as their interests because skills such as ``Web Development'' can also be an interest, and people usually are good at things that they are interested in. For college students, we extracted hashtags (excluding prefix ``\#'') as interests. A hashtag is a user-defined, specially designated word in a tweet, prefixed with a ``\#'' [31]. Originally, we experimented LDA topic modeling to discover topics of interests from all college students' tweets and intended to use these to define each student's interests. However, due to the noise and non-interest related terms in tweets (excluding stop words and non-English words), most of the terms generated are too generic to be defined as topics of interests. Therefore, we extracted one's unique hashtags as proxy for interests. Hashtags have been used in characterizing topics in tweets [23] and have shown to be interest-related to a decent extent [32]. Although high-frequency hashtags are intuitively better representations of one's interests, including all unique hashtags allows us to extract a wilder range of interests. After we extracted interests from both students and role models, we stored everyone's interests as a set which we call \emph{interest set}. The size of the set varies from user to user depending on the number of interests of that user. 

\subsection{Ranking Algorithm}
We rank all STEM role models for each student based on the similarities of their attributes. Specifically, for each comparison of a student and a role model, we calculate the similarity of each attribute, and rank the role model based on the arithmetic average across similarities of all attributes. We will now explain our methods used for each comparison.

For gender and race, we simply compared if the two people have the same string for gender or race. In our case, there are two strings for gender, ``female'' and ``male'', and five strings for race, ``White'', ``Black'', ``Asian'', ``Api'', and ``Hispanic''. Therefore, the \emph{gender similarity} is either 1 or 0 because two people either have the same gender or not, and the same went for \emph{race similarity}. 

For geographic locations, we used string comparison method to measure the similarity of two locations. Originally, we experimented two ways to calculate it: the actual distance between two locations based on their latitudes and longitudes, and the Levenshtein distance between the two strings that represent the two locations. Due the variety of possible expressions of the same location, traditional tool such as geocoder \footnote{\url {https://github.com/geopy/geopy}} can only correctly convert well-formatted locations that do not contain non-letter characters. For example, a real college student has location ``buffalo state '18 psych majorr'' and it cannot be successfully converted into coordinates using geocoder, but clearly that the student studies in buffalo. Since our objective is to be able to compare as many locations as possible, we decided to use string comparison, which allows the flexibility of using various location representations for the same place. Specifically, we employed Levenshtein distance \footnote{Levenshtein distance is the minimum number of single-character edits required to change one string into the other, and it is applicable to strings with different lengths. \url{https://github.com/miohtama/python-Levenshtein}} [13] to calculate the distance between two strings, and the \emph{Levenshtein-based similarity} (a ratio between 0 and 1) is defined as:
\begin{equation}
   \frac{\emph{length}(S_1)+ \emph{length}(S_2) - \emph{Levenshtein distance}(S_1,S_2)}{\emph{length}(S_1)+ \emph{length}(S_2)}
\end{equation}
where in the case of location, \begin{math}S_1 (S_2)\end{math} is the string of the student's (role model's) location, and we then have our \emph{location similarity}. A minor problem of this similarity measure is that two geographically different locations might contain similar words and have a high similarity, such as ``Washington D.C.'' and ``Washington State''. But this happens relatively rare only if there are enough people from one of the location or both.

We used Jaccard coefficient [10] combined with \emph {Levenshtein-based similarity} to compute the similarity of two \emph{interest sets}. Hashtags are often not real words but a combination of words without spaces. While a real student's hashtag ``computersciencelife'' and a real role model's interest ``computer science'' clearly refer to the same interest in the field of computer science, the two strings are different and have a \emph{Levenshtein-based similarity} of 0.86. Therefore, in order to capture the overlapping interests between two \emph{interest sets}, we need a threshold for \emph{Levenshtein-based similarity} that decides whether two strings refer to the same interest. After extensive experimenting with real data, we chose our threshold to be 0.8. Our \emph {interest similarity} is then defined as:
\begin{equation}
   \frac{\vert{I_1}\bigcap{I_2}\vert} {\vert{I_1}\bigcup{I_2}\vert} = 	\frac{\text{\# of overlapping interests}}{\vert{I_1}\vert + \vert{I_2}\vert - \text{\# of overlapping interests}} 
\end{equation}
where \begin{math}I_1(I_2)\end{math} is the student's (role model's) \emph{interest set}. A potential problem is that since our measurement is string-based but not concept-based, it might not capture the synonymous of interests as overlapping interests. 

After we calculated the similarities of all four attributes, we combined them by taking the arithmetic average and used that to rank the role models. In the cases of missing values, any unlabeled attributes is not taken into account. For instance, if a student does not have gender information, the arithmetic average will entirely depend on the similarities of his other three attributes.

\subsection{Evaluation}
In this section, we verified our ranking algorithm on 2,000 college students from the 297 most populated cities\footnote{with a population of at least 100,000} in the United States [28]. All these students are randomly selected from our database. We manually evaluated their top-5 ranked role models, and we also recommended these role models to them via Twitter.

Although it is desirable to evaluate the ultimate impact of our study, we recognize that this would require tracking the subjects of the study over their career of substantial length (e.g., over 10 years). Therefore, it is beyond the scope of this study, and we decided to use matching accuracy as the performance measure, which is defined as:
\begin{equation}
   \frac{\text{\# of students that were \emph{correctly} matched with n role models(s)}}{\text{\# of total students}}
\end{equation}
where n is the second metric, the specific number of role models out of the top-5 that are correctly matched with the student. It represents the granularity level of matching. We consider a student is \emph{correctly} matched with a role model if the LinkedIn user is indeed a STEM role model and has the same gender, race and geographic location as the student. We did not evaluate interests since they are often not explicitly stated in social media and it would be too difficult to discover every student's real interests by reading his tweets.

We took a careful effort to manually evaluate the matching results of these 2,000 college students by checking their Twitter profile pages and the LinkedIn profile pages of their top-5 ranked STEM role models. We utilized all the information on their respective social media profiles to determine their gender, race, and geographic location. In order to determine if someone is indeed a STEM role model, we make our best judgment, as a career counselor would, based on the entire LinkedIn profile, which usually includes demographic background, personal summary, industry, education, working experience and skills.

If we failed to determine any of the three attributes of a student, we will have to consider that he is not correctly matched with any role model because we are unable to conduct the evaluation. Consequently, for Twitter public accounts and students with unlabeled gender, race, or invalid location, they all receive zero correctly matched role models. Location should not have been a problem since we selected these students by their locations, but we found that a handful of students have removed or changed their locations after we collected the data.
\begin{table}
\vskip -0.15in
\caption{Top: top-5 role models for a White, male student from ``Atlanta, Georgia''; Bottom: top-5 role models for an Asian, female student from ``Round Rock, TX''}
\centering
\includegraphics[height=6.2cm,width=\linewidth,keepaspectratio]{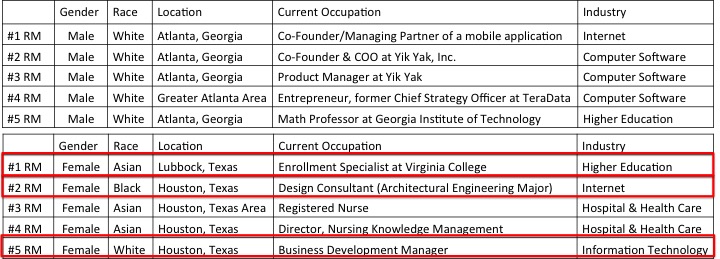}
\vskip -0.1in
\end{table}

Table 1 shows two randomly selected representative examples of the matching results for two students. We consider that the student in the top table was correctly matched with all five role models and the student in the bottom table was only correctly matched with \#3 and \#4 role models at state-level because \#1 role model is not in STEM-related occupation and \#2 and \#5 are not Asian. None of the role models was correctly matched at city-level.

Taking into consideration that our limited database of STEM role models may have an impact on the performance of the ranking algorithm, we conducted evaluation in four levels: city-level for 297 cities, state-level for top 297 cities, city-level for top-10 cities, and state-level for top-10 cities. Among the 2,000 selected students, about a quarter of the selected students are from the top 10 cities. Intuitively, we expect more students to be correctly matched with role models at the state-level than city-level. Also, we expect students from the top-10 cities to be correctly matched with more STEM role models because there should be more diverse role models in these cities.
\begin{figure}
\vskip -0.1in
\centering
\includegraphics[height=6.2cm]{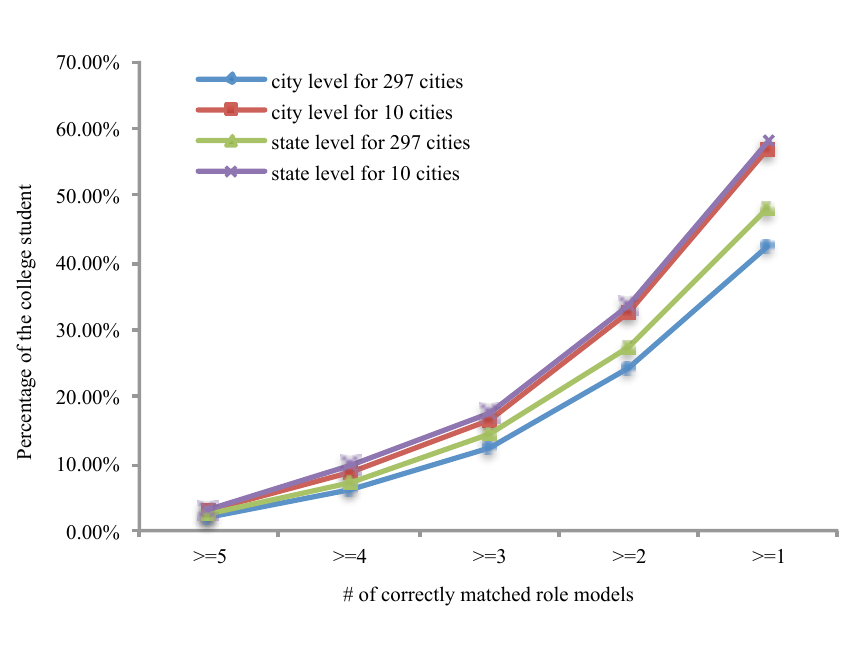}
\vskip -0.1in
\caption{The results of matching accuracy of 2,000 students from 297 cities in four different levels.}
\vskip -0.1in
\end{figure}

In Fig.6 we show the overall matching accuracy in the four levels. We first look at our baseline, the city-level for 297 cities. 42\% of the college students were correctly matched with at least one role model. We noticed that around half of them was not matched with any role model and this is partly due to those college students with unlabeled gender and race information. 

We then noticed that our ranking algorithm performs better in the 10 cities than in the 297 cities for both city and state levels. Numerically, the difference increases as the the minimum number of correctly matched role models decreases. If we look at students who were correctly matched with at least one role model, for both city and state levels, the top-10 cities outperforms the 297 cities by 33\% and 21\%, respectively; the ranking algorithm achieves a decent accuracy of 57\% in both city and state levels for the 10 cities. Also, our ranking algorithm performs better in the state-level than in city-level for the 297 cities. With students who were at least correctly matched with one role model, the difference is 13\%, which is smaller but still very significant. However, there is almost no difference in state and city levels for the top-10 cities. A possible explanation is that because there are more STEM role models of various types in the top-10 cities, the student can usually get matched with STEM role models who are from the exact same city. 

During our evaluation, we are encouraged to see that there is a good variety of STEM role models in different industries even for students with the same demographic background. We think this is a positive indicator that the attribute, interests, in fact contributes to our ranking algorithm.
\begin{figure}
\centering
\includegraphics[height=6.2cm,width=\linewidth,keepaspectratio]{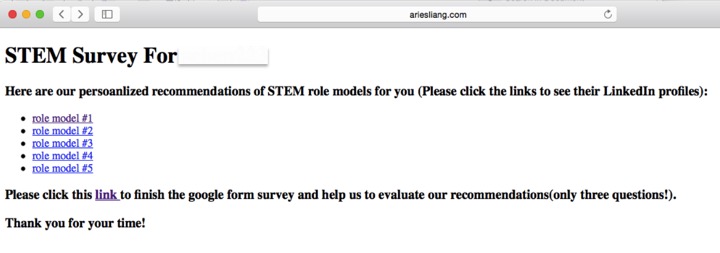}
\vskip -0.1in
\caption{An example of the personalized webpage for a real college student user on Twitter.}
\vskip -0.1in
\end{figure}

In order to make a real-life impact, for each student, we generated a personalized webpage and delivered the link of the webpage via tweeting at him from the official Twitter account of our study. Fig.7 shows an example of such webpage. It contains the LinkedIn public profile links of his top-5 role models and a survey regarding the accuracy of our recommendations. We only received a very small number of responses and conducted preliminary analysis. All responses indicated that they are indeed currently college students, a third agree that the recommendations are good and a third indicated that they would be more interested in STEM majors/careers if they had role models in STEM fields. We would need more responses to validate our implementation, and a potential way to do so is to cooperate with our university, apply the ranking algorithms on students who are Twitter users and ask for responses.

\section{Conclusion And Future Work}

In this paper, we present an innovative social media-based approach to promote STEM education by matching college students on Twitter with STEM role models from LinkedIn. Our ranking algorithm achieves a decent accuracy of 57\% in the city-level for the top-10 cities that the STEM role models come from. We also design a novel implementation that recommends the matched role models to the students. To achieve this, we identified college students from Twitter and STEM role models from LinkedIn, extracted race, gender, geographic location and interests from their social media profiles, and developed a ranking algorithm to rank the top-5 ranked STEM role models for each student. We then created a personalized webpage with the student' role models and recommended the webpage to the student via Twitter. 

Our recommendation is not imposed on either side. It is the students' choice if they want to initiate the connection with the role models via LinkedIn or other methods; and it is for the role models to decide if they want to accept their LinkedIn invitations or other forms of communication. In the case of LinkedIn, note that if a student decides to approaches a potential role model, he can express why he would like to get connected (e.g., interest in STEM fields), and the role model can make his own judgment. One may worry that our implementation of recommendations may be considered a form of spamming on students, however, our intention is clearly to help their careers, and \emph{not to profit} from them. 

There are several possible extensions of our study in the future. Our approach might have a reduced effect for college seniors since it is more difficult for them to switch majors. However, it is not uncommon that students change their career paths after graduation, and in the future we could recommend role models with similar experiences specifically to seniors. We could also expand our target population to high school students or focus on promoting STEM education specifically to minority college students. In addition, we could classify STEM role models into specific groups such as current STEM major college students and experienced STEM role models since students might feel more comfortable reaching out to their peers. Finally, we could design an application based on our implementation to achieve real-time matching, where a college student could log into our application using their Twitter account, and we could collect their data, extract their attributes, and give them STEM role model recommendations in real-time. This application could also be generalized to other social media since many methods we used are compatible with other platforms.

We hope this study can serve as a starting point to make use of the rich data and powerful networking ability of social media ``by the people'' in order to promote STEM education and build positive influence ``for the people''.

\subsubsection*{Acknowledgments.} 

This work was supported in part by Xerox Foundation, and New York State through the Goergen Institute for Data Science at the University of Rochester. We thank all anonymous subjects for contributing to the evaluation of our system. 

\label{references}




\end{document}